\documentclass[a4paper,11pt]{article}
\usepackage{jheppub} % for details on the use of the package, please
\usepackage[T1]{fontenc} % if needed

\title{\boldmath Non-Extensive Behaviour of the QCD Strong Coupling Constant}

\author[a]{K. Javidan,}
\author[b]{M. M. Yazdanpanah,}
\author[b]{and H. Nematollahi}

\affiliation[a]{Department of Physics, Ferdowsi University of Mashhad, 91775-1436 Mashhad, Iran}
\affiliation[b]{Faculty of Physics, Shahid Bahonar University of Kerman, Kerman, Iran}

\emailAdd{javidan@um.ac.ir}
\emailAdd{myazdan@uk.ac.ir}
\emailAdd{hnematollahi@uk.ac.ir}

\abstract{The compatibility of theoretically calculated values for $\alpha_s(Q)$, through the renormalization group approach with experimental data is studied. There exists considerable divergence in-between theoretical and experimental results at low energies, which cannot be explained by thermal field theory and considering chemical potential. Such great deviation can be treated successfully by considering the q-generalized statistical effects through adding a q-nonextensive parameter in the fitting of theoretical results on the experimental data. }

\begin{document} 
\maketitle
\flushbottom

\section{ Introduction}
\label{sec:intro}

Fundamental parameters of quantum chromodynamics (QCD) theory are the strong coupling $g_s$ (or $\alpha_s=\frac{g_s^2}{4\pi}$ ), quark masses $m_q$ and some additional degrees of freedom (like CP-violating terms). The strong coupling $\alpha_s$ is one of the three fundamental coupling constants of the standard model (SM) of particle physics, which is related to the colour part ($SU(3)_C$) of the SM gauge symmetry. While $\alpha_s$ is called a constant, the strength of strong coupling (actually all couplings of the SM) is a function of the energy scale (or momentum transfer) $Q^2$ of particular process under consideration. The important property of the QCD which is known as asymptotic freedom, originates from this fact that $\alpha_s(Q^2)$ decreases by increasing the transfered energy $Q^2$. Measuring (or accurately calculating) the strong coupling in both high and low energy scales, provides us important knowledge about the nature of strong force as a fundamental interaction. On the other hand, at low energies running coupling constant plays an important role in the binding of quarks and gluons together into nucleons. It is obvious that, we need the most accurate value of the $\alpha_s(Q^2)$ at each quantity of transferred energy that involves the strong interaction to compare theoretically predicted results with experimental outcomes \cite{Prosperi:2006hx,Deur:2016tte}.

Considering ultraviolet divergences due to loop diagrams in Feynman graphs leads to the renormalised (running) coupling constant $\alpha_s$, as a function of the renormalization scale $\mu_0^2$. If we choose $\mu_0^2$ around the scale of the momentum transfer $Q^2$ in a desired strong process, then $\alpha_s(Q^2)$ can be considered as the effective strength of the strong  interaction at that process. If $\alpha_s(\mu_0^2)$ at scale $\mu_0^2$ cannot be determined, we can calculate it using a renormalization group differential equation containing several functions at different orders of $\mu_0^2$. Thus, its solution contains an extra constant of integration, which should be fixed from experimental data. The constant of integration is mostly fixed in the energy scale of $Q^2\approx M_Z^2$, in which $M_Z$ is the mass of $Z$ boson. All needed calculations can be done using quantum field theory (QFT) at zero temperature as well as QFT at finite temperature \cite{Kapustabook}. There are many works on solving this equation with different orders of renormalization functions at zero and non-zero temperatures.  
When we speak about non-zero QFT, we must consider a thermodynamically described system in equilibrium or out of its equilibrium state indeed. Application of thermodynamics approach to explain some features of QCD interactions is not new \cite{tpdf1, tpdf2}. Thermodynamical model for proton spin \cite{tpdf1} and parton distribution functions (PDFs) are bright examples of such investigations \cite {tqcd0}, beside its applications to explain other features of QCD interactions \cite{tqcd1, tqcd2}. A well-known application of thermodynamics is using the Fermi-Dirac distribution for quark and antiquark partons and Bose-Einstein distribution for gluons in nucleons. It results a fair description for the x-dependence of PDFs \cite {tqcd0}.\ 

In recent years several research works have been done based on the Tsallis' formalism  \cite{Tsallis:1987eu,Tsallis:1998ws,Tsallis:book}. Some deviations of theoretical predictions based on Maxwell-Boltzmann distributions, from experimental data can also be treated by considering long range interactions which leads to Tsallis non-extensive statistical model. Investigations show that nonextensivity is considerable when long range interactions and non-perturbative effects are taken into account. As an example in high energy physics applications, it is shown that considering nonextensivity can successfully explain experimental distribution of transverse momentum of hadrons with respect to the jet axis ($p_T$) in $e^+e^- \rightarrow hadrons$ reaction \cite{ee}. The effects of nonperturbative long range interactions in $p-p$ collision has also been studied using the LHC experimental data \cite{pp1, pp2}. The results of applying Tsallis statistics generally improves predicted properties of hadronic systems in most of the cases \cite{had1, had2}. In some studies the non-extensive generalization has been applied to investigate the q-generalized Bose-Einstein and Fermi-Dirac distributions of many particle systems \cite{Teweldeberhan:2005wq,Conroy:2010wt,Silva:2009xra,mitra}. Motivated by above studies, we investigate the effects of considering non-extensive statistics on the prediction of running coupling of strong interaction $\alpha_s(Q^2)$, especially at small values of $Q^2$, where theoretical calculations are drastically far from the experimental data.

The outlines of this paper are as follows: in section 2 we briefly describe the properties of the strong running coupling constant and present the results in two different scenarios: in subsection 2.1 the behaviour of
 $\alpha_s$ is studied in the standard perturbative QCD framework while its properties in the thermal perturbative QCD is probed in the subsection 2.2. We investigate the role of q-generalized non-extensive statistical effects on the properties of $\alpha_s$ in section 3 and present its results in this section too. We render the conclusions and remarks in section 4.

\section{ Running of the QCD coupling constant}
\label{sec:RQCC}
\subsection{$\alpha_s$ in Standard Perturbative QCD}

The strength of QCD as a non-Abelian gauge theory of the strong interactions, is evaluated by the strong coupling $\alpha_s$. 
Variation of strong coupling respect to the normalization scale $Q$ is given by the renormalization group equation as follows:

\begin{equation}\label{beta}
Q^2 \frac{d}{d Q^2} a_s(Q)=\beta a_s=-\sum_{i=0}^{N}\beta_i a_s^{i+2}
\end{equation}
where $a_s=\frac{\alpha_s}{\pi}$, $N$ is the number of loops involved in the calculation and $Q$ is the normalization scale of renormalization group approach\cite{Peterman:1978tb,be2}.  
At the order of $N=0$, the equation (\ref{beta}) has an exact solution:
\begin{equation}\label{beta0 sol}
a_s(Q)=\frac{1}{\frac{1}{a_s(\mu_0)}+2\beta_0 \ln \left( \frac{Q}{\mu_0} \right) }
\end{equation}
where $a_s(\mu_0)$ is the strong coupling in the normalization scale $\mu_0$,  $\beta_0=\frac{1}{4}\left(11-\frac{2}{3}n_f\right)$ and $n_f$ is the number of active flavours in the energy scale \cite{5loop}. As maintained before, $\mu_0$ is taken generally as $\mu_0=M_Z$ while $M_Z$ is the Z boson mass. The world average value for this parameter is $\alpha_s(\mu_0)=\pi a_s(\mu_0)=0.1181 \pm 0.0011$ as reported in the Ref. \cite{alphas2019}. By calculating the higher order QCD corrections, the perturbative coefficients of functions $\beta_i$ have been obtained at different loop levels \cite{Politzer:1973fx,Gross:1973id,Caswell:1974gg,Avdeev:1980bh}.  
From our best of knowledge, the QCD $\beta$-functions have been calculated for up to five-loops, using the zero temperature QFT \cite{5loop}. $\beta$-functions for the four-loop calculations are also available \cite{4loop} and they are in agreement with each other. We have generate the theortical values of $\alpha_{s}(Q)$, at one-, three- four- and five-loop calculations. Figure 1 presents one-loop, three-loop and five-loop running coupling as well as available experimental data. For energies below $Q=3 GeV$ active quark flavours has been taken as $n_f=3$. For $3 GeV\le Q \le 10 GeV$, we set $n_f=4$ and for $Q > 10 GeV$ number of active quarks has been set as $n_f=5$.  Experimental data have been collected from the CMS \cite{cms1,cms2, cms3, cms4, cms5}, D0 \cite{d01, d02} and ATLAS \cite {AT} collaborations and particle data group report \cite{pdg}. The most important data in our calculations are strong coupling at low energies which are collected from the H1 collaboration \cite{H11,H12}.    \

\begin{figure}[htp]\label{f1}
\centerline{\begin{tabular}{cc}
\includegraphics[width=12 cm, height= 10 cm]{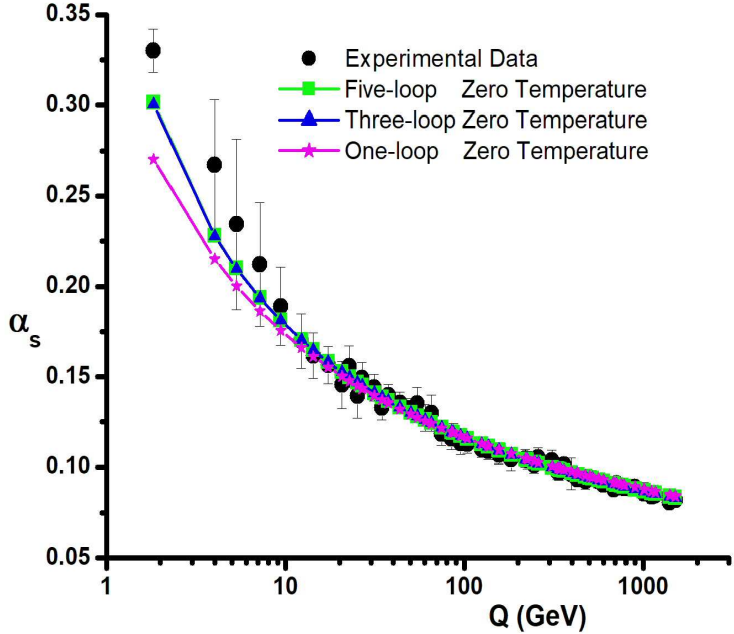}
\end{tabular}}
 \caption{\footnotesize
Running of the strong coupling calculated by one-, three- and five-loop renormalization using QFT at zero temperature. }
\end{figure}
We have calculated the variance as:
\begin {equation}
\chi^2=\sum_{i} \left(\frac{\alpha_s^{Th}-\alpha_s^{Ex}}{\delta}\right)^2
\end {equation}
where $\alpha_s^{Th}$, $\alpha_s^{Ex}$ and $\delta$ are "calculated value", "experimental value" and the "total error of measuring $\alpha_s(Q)$", respectively.  
Indeed, we use above definition for $\chi^2$ just for comparing results of different calculations. The variance $\chi^2$ for one-loop calculation is obtained $\chi^2=54.337$ and in three-loop approximation it is $\chi^2=24.299$, while for the five-loop calculations it becomes $\chi^2=23.782$. Comparing the values of $\chi^2$  at one-, three- and five-loop approximations shows that we would not expect to find magic results at more-loop approximations. The results of our calculations are in agreement with similar published outcomes. For example one can compare figure 1 with figure 12 of Ref. \cite{CMS10}. Figure 1 clearly indicates that, QCD calculations for running of the strong coupling need some treatments especially at low energies.

 \subsection{$\alpha_s$ in Thermal Perturbative QCD}
When we  study the particle distribution function, we cannot ignore thermodynamics based on a kind of statistical kinetic theory. Such model can be constructed by regarding the resummation on individual characteristics of particles under consideration. Formal field theory is constructed for the systems in equilibrium states at zero temperature. Indeed, by replacing the vacuum expectation values with quantum statistic expectation values, field theory for finite temperature is obtained \cite{tft,htl}. Actually, We expect that the thermal field theory provides more realistic results for the dynamics of the QCD interactions. Using results of thermal field theory, we can include effects of chemical potential as well as temperature considerations into our calculation. It is clear that,  such effects are considerable in low energies, where purtuarbative and long range effects become important. Fortunately, two- and three-loop QCD gauge coupling have been calculated in the context of thermal field theory, but with some approximations \cite {2htl, 3htl,Andersen:2015eoa}.
The two-loop effective QCD gauge coupling at the temperature T is calculated as \cite {2htl}:
\begin{equation}\label{2loop}
g^2 (T,{\bar \Lambda})=T\left\{g^2({\bar \Lambda})+\frac{g^4({\bar \Lambda})}{\left(4\pi\right)^2}\alpha_{E7}+O(g^6)\right\}
\end{equation}
where 
\begin{equation}\label{E7}
\alpha_{E7}=-\beta^\prime_0 \ln \left( \frac{{\bar \Lambda} e^{\gamma_0}}{4 \pi T}\right)+\frac{1}{3}C_A-\frac{16}{3}T_F \ln(2)
\end{equation}
while $\gamma_0$ is the Euler gamma-constant, $C_A=N_c$, $C_F=\frac{N_c^2-1}{2N_c}$, $T_F=\frac{N_f}{2}$ (we take $N_c=3$, $N_f=3$ ) and the normalization scale parameter is denoted by ${\bar \Lambda}$. The ${\bar \Lambda}$ acts as normalization scale $Q$ in (\ref{beta}). The $g({\bar \Lambda})$ is calculated from the renormalization group equation:
\begin{equation}\label{1run}
{\bar \Lambda} \frac{d}{d{\bar \Lambda}}g^2({\bar \Lambda})=\frac{\beta^\prime_0}{(4\pi)^2}g^4({\bar \Lambda})+ \frac{\beta^\prime_1}{(4\pi)^4}g^6 ({\bar \Lambda})+O(g^8)
\end{equation}
in which 
\begin{eqnarray}\label{2run}
\beta^\prime_0&=&\frac{-22C_A+8T_F}{3}\nonumber \\
\beta^\prime_1&=&\frac{-68C_A^2+40C_AT_F+24C_FT_F}{3}.\\ \nonumber
\end{eqnarray}
Simply one can find that $\beta_0$ in (\ref{beta0 sol}) is equal to $\beta^{\prime}_0$ in (\ref{2run}). It should be noted that, effective coupling (\ref{2loop}) has been presented up to order $O(g^8)$ in \cite{2htl}. Higher orders is needed, if we consider $O(g^6T^4)$ contribution to the pressure of hot QCD which is negligible in this work.

The three-loop running coupling, approximately can be calculated as: \cite{3htl,Andersen:2015eoa,3htl ref}:
\begin{eqnarray}\label{3loop}
\alpha_s(\bar \Lambda)=&&\frac{\pi}{b_0t}[1-\frac{b_1}{b_2}\frac{\ln t}{t} +\frac{b_1^2((\ln t)^2-\ln t -1)+b_0b_2}{b_0^4t^2}\nonumber\\&&
-\frac{b_1^3((\ln t)^3-\frac{5}{2}(\ln t)^2-2 \ln t+\frac{1}{2})+3b_0b_1b_2 \ln t}{b_0^6t^3}]
\end{eqnarray}
where $t=\ln \frac{\bar \Lambda^2}{\Lambda_{\overline{MS}}^2}$ and
\begin{eqnarray}\label{3loopq}
b_0=&&\frac{11C_A-2N_f}{12 \pi}\nonumber\\ 
b_1=&& \frac{17C_A^2-5C_AN_f-2C_fN_f}{24\pi^2}\\
b_2=&&\frac{2857C_A^3+\left(54C_f^2-615C_fC_A-1415c_A^2 \right)N_f+\left(66C_f+79C_A \right)N_f^2}{3456\pi^3}.\nonumber
\end{eqnarray}
where the $\bar \Lambda$ has been taken (at the rest frame) as follows: \cite{3htl} :
\begin {equation} \label{lam}
{\bar \Lambda_0}=2\pi\sqrt{T^2+\mu^2/\pi^2} .
\end {equation}
It is clear that, in the center of mass frame, we should add the energy $Q$ to the thermal energy and the chemical potential.
The $\Lambda_{\overline{MS}}$ can be found with a suitable condition, using available experimental data. We can set the $\Lambda_{\overline{MS}}$ with $\alpha_s$ at low energies or typically at the energy of the $M_Z$. According to \cite{3htl,Andersen:2015eoa} we set $\Lambda_{\overline{MS}}$ with condition
$\alpha_s(1.5 GeV)=0.326$, in order to compare our results with outcomes of \cite{2htl} and \cite{3htl}. Our result was $\Lambda_{\overline{MS}}=337 \pm 8 MeV$ in three loop approximation and 
$\Lambda_{\overline{MS}}=240 MeV$ at two loop calculation, which is a slightly different from what has been reported in \cite{3htl} as
 $\Lambda_{\overline{MS}}=316 MeV$ for three loop calculation. Our results are based on different experimental data set, from what have been used in \cite{3htl}. It may be noted that, we have collected the newest available data and also we have considered experimental data at low energies too.

In order to evaluate the strong coupling using the finite temperature approach from Eqs.(\ref{2loop}) and (\ref{3loop}), we need an estimation for the temperature. It is clear that, at low energies, the effects of temperature in the value of strong coupling are considerable as one can find from Eq.(\ref{lam}). Acceptable estimation of temperature at low energies can be borrowed from the calculation process of parton distribution functions using the thermodynamical model for proton \cite{temp1, temp2, temp3}. Table 1 contains results of fits to ABCLOS and CDHS data on parton distribution function of proton \cite{temp2}. We have calculated $\alpha_s(Q)$ at two- and three-hard thermal loop perturbation approximation and compared the results with available experimental data. In the figure 2 we present the results of three-loop evaluation of running $\alpha_s(Q)$ obtained using evolution equation (\ref{3loop}). Total variance $\chi^2$ for these results is $\chi^2=45.147$ which is not better than outcomes of zero temperature. It is because of using an imprecise value as initial for $\Lambda_{\overline{MS}}$ condition at low energies with a very large uncertainty in comparison with a more accurate value of $\alpha_s{(M_Z)}$. In order to find the best initial condition, we have set up global fits to find the best value of $\alpha_s(M_Z)$ which gives lowest value for the $\chi^2$ at one-, three- and five-loop  formal QFT calculations as well as three-loop thermal calculation. Table (2) shows results of best values for $\alpha_s{(M_Z)}$ which minimize the $\chi^2$ and also related minimum value of $\chi^2$ in zero temperature as well as finite temperature calculation. However, the thermal field theory gives a slightly better value for the $\alpha_s(M_Z)$, but considering the $\chi^2$ indicates that differences are not meaningful. Anyway, according to the contents of the table 2, our global fit of theoretical models on experimental data clearly show that, both formal QFT and finite temperature QFT (globally) provide almost same results on all range of experimental data.\

Figure 3 demonstrates results of calculated $\alpha_s$ using three-loop thermal field theory with $\alpha_s(M_Z)=0.11805$. Comparing figures 2 and 3 clearly show that, if we adjust theoretical and experimental values at higher energies, then we lose validity at lower energies and vice versa. It is found clearly from figures 1, 2 and 3 that running of strong coupling constant calculated at zero temperature and finite temperature fail to fit on low energy experimental data, where long range effects of QCD interactions are not negligible. On the other hand, figures 1-3 show that, we would not expect a great change in the results, by employing higher orders of perturbative calculations. In the next section we present a non-extensive treatment for running of the strong coupling based on the thermal field theory.

\begin{table} \label {t1}
\begin{center}
\begin{tabular}{ c   c   c }
\hline \hline
$ Q (GeV) $ &    T (MeV) &  $\delta T$ \\ \hline
$ 2.5 $ &	 55 & $\pm$ 5\\
$ 4 $ &	58 & $\pm$ 8\\
$ 7 $&	54 & $\pm$ 6 \\
$ 20 $ & 51 	& $\pm$ 5\\
$ 40 $&	51 & $\pm$ 2\\
$ 60 $&	51 & $\pm$ 2\\
$ 80 $ & 47.6 & $\pm$ 0.5\\\hline
\end{tabular}
\end{center}
\caption{Results of fits to ABCLOS and CDHS data on parton distribution function of proton, presented in \cite{temp2}. }
\end{table}

\begin{table} \label {t2}
\begin{center}
\begin{tabular}{ c   c   c }
\hline \hline
Order &   $ \chi^2_{min}  $ &  $\alpha_s(M_Z)$ \\ \hline
one-loop &	 46.311 & 0.11850\\
three-loop &	23.320 & 0.11800\\
five-loop&	21.850 & 0.11800 \\\hline
3-loop thermal & 27.903 & 0.11805\\ \hline
\end{tabular}
\end{center}
\caption{The best values for $\alpha_s(M_Z)$ which minimize $\chi^2$ in one-, three-, and five-loop zero temperature calculations and three loop thermal field theory. }
\end{table}

\begin{figure}[htp]\label{f2}
\centerline{\begin{tabular}{cc}
\includegraphics[width=12 cm, height= 10 cm]{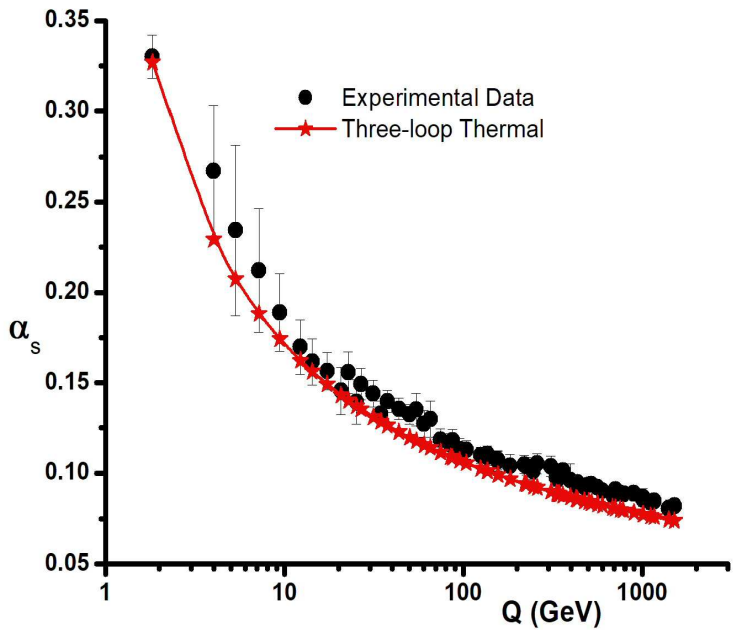}
\end{tabular}}
 \caption{\footnotesize
Running of the strong coupling calculated by three-loop renormalization using thermal QFT with initial condition $\alpha_s(1.5 GeV)=0.326$. }
\end{figure}

\begin{figure}[htp]\label{f3}
\centerline{\begin{tabular}{cc}
\includegraphics[width=13 cm, height= 10 cm]{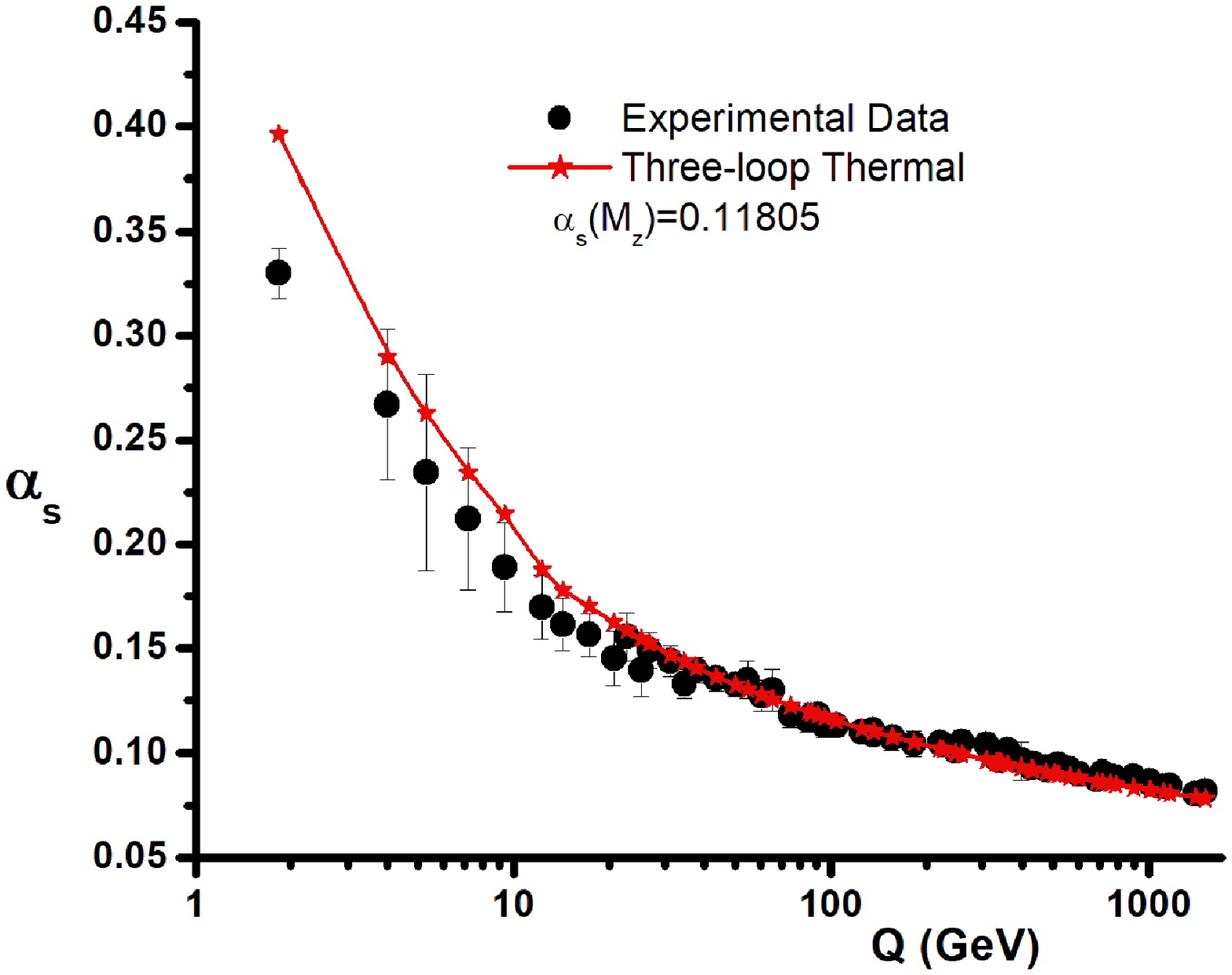}
\end{tabular}}
 \caption{\footnotesize
The result of three-loop thermal field theory for running coupling $\alpha_s$ with best value for minimizing the $\chi^2$ as $\alpha_s(M_Z)=0.11805$. }
\end{figure}

\section{ q-Generalized QCD Running Coupling Constant}
\label{sec:RSCE}

A thermodynamical model for the nucleon has been constructed by considering the valence quarks as Fermi-Dirac distributed noninteracting particles which are bounded in a bag of confining volume V at a certain temperature $T$ \cite{tpdf1, tpdf2}. This model is able to explain some features of deep inelastic scattering (DIS) with acceptable approximation. The prediction of parton distribution function is an interesting obtained result by this model, however the results need some treatments \cite{buch}. The Boltzmann-Gibbs formulations cannot explain all features of a system  correctly, if the system contains sources of fluctuations (for example in its temperature and/or number density ... ) or if there exists long-range correlation in the dynamical process. It is shown that a modified thermodynamical model based on the q-generalized Bose-Einstein and Fermi-Dirac systems provide more accurate results \cite {mitra, qn}. Within the context of q-generalized statistical models, the effective coupling for a non-extensive system in the rest frame, approximately is presented as following \cite{mitra}:
\begin {equation} \label{main}
\alpha_s(T,\bar{\Lambda},q)=\alpha_s(T,\bar{\Lambda})\left[1+(q-1)\frac{\frac{2N_c}{\pi^2}\zeta(3)-\frac{2N_f}{\pi^2}Polylog\left[3,-1\right]}{\frac{N_c}{3}+\frac{N_f}{6}+\frac{\bar{\Lambda}^2N_f}{2\pi^2T^2}}   \right]
\end {equation}
where $q$ is nonextensive parameter, $\zeta(n)$ is the Riemann zeta function and 
$Polylog \left[n,-1\right]=\sum_{k=1}^{\infty}\frac{(-1)^k}{k^n}$. 
This equation changes the running of coupling, especially in low values of $Q$, where chemical potential and thermal energy is noticeable. One can find a suitable value for the nonextensive parameter $q$ to fit experimental data on theoretical model. 
As variation of temperature (in comparison with statistical error) is not significant (please see the table 1), we took fixed value for temperature as $T=51 MeV$. Running of gauge coupling at two-loop approximation of thermal QCD (\ref{2loop}) can be written as $\alpha_s(T,\bar{\Lambda})=T\alpha_s(\bar{\Lambda})$. Thus we can easily use equation (\ref{2loop}) in (\ref{main}) to calculate q non-extensive treatment of the gauge coupling. Figure 4 demonstrates results of q-nonextensive fit of two-loop thermal calculation on experimental data. Minimum uncertainty for q-nonextensive fit is $\chi^2=12.24$ which obtains with $q=1.2412$. Reduction of deviation from experimental data at low energies in a q-nonextensive regime is a very interesting outcome.

\begin{figure}[htp]\label{f4}
\centerline{\begin{tabular}{cc}
\includegraphics[width=13 cm, height= 10 cm]{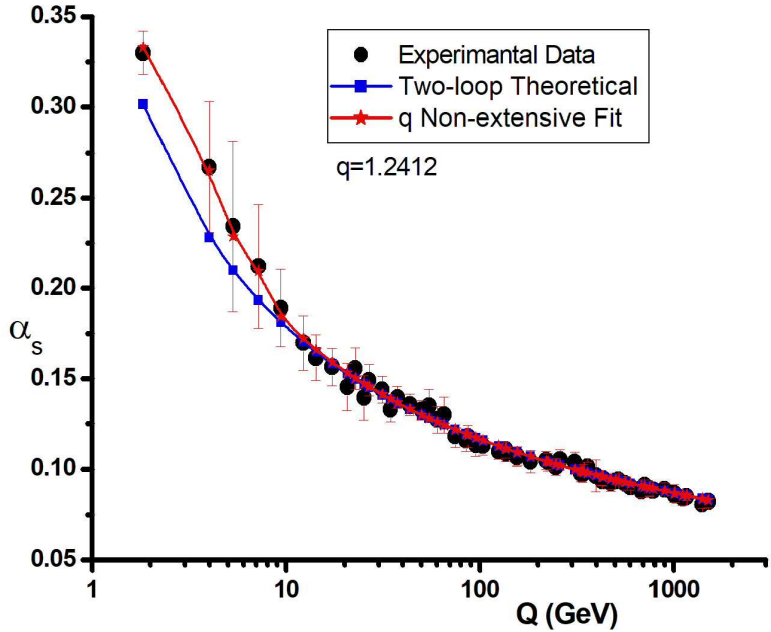}
\end{tabular}}
 \caption{\footnotesize
The result of $\alpha_s$ calculated using two-loop thermal field theory and results of q-nonextensive fit based on two-loop thermal field calculation on experimental data. }
\end{figure}

Calculation of q non-extensive gauge coupling in three-loop formulation is more complicated as it appears as $\alpha_s(\bar{\Lambda})$. 
Contributions of thermal energy and chemical potential have been mixed in $\bar{\Lambda}$, where in the rest frame it is identified by (\ref{lam}). In the centre of mass frame, we must add $Q$ to this quantity.  Fortunately, thermal energy is very smaller than chemical potential (and also $Q$ in the centre of mass frame). Thus, as an excess approximation, we have used a constant temperature in our calculations. Figure 5 shows results of numerical optimization over the q value, which is obtained with $q=1.3181$. Maximum error in this calculation is $\chi^2=11.93$ which is not very better than results of the two-loop thermal approximation.

\begin{figure}[htp]\label{f5}
\centerline{\begin{tabular}{cc}
\includegraphics[width=13 cm, height= 10 cm]{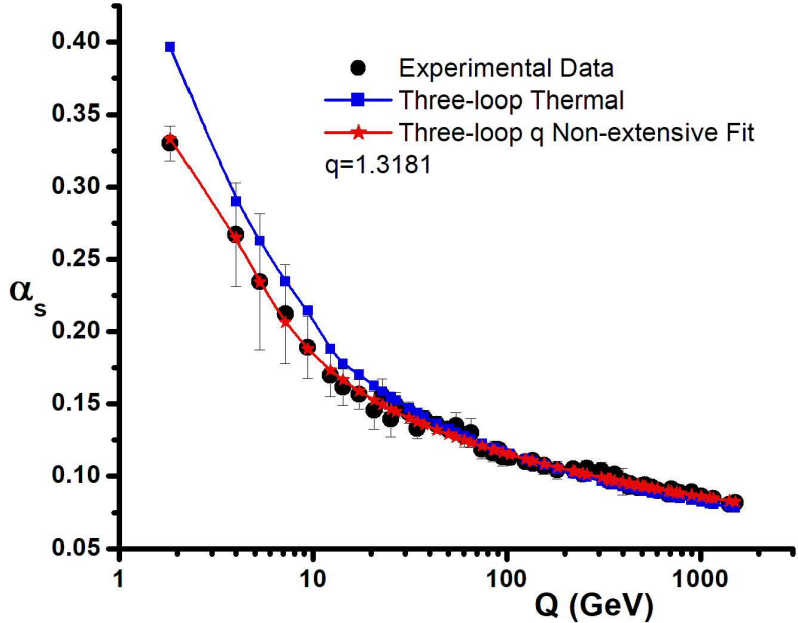}
\end{tabular}}
 \caption{\footnotesize
 Calculated $\alpha_s$ using three-loop thermal field theory, q-nonextensive treated values and experimental data. }
\end{figure}

\section{ Conclusions and remarks } 
\label{sec:Con}
One-, three- and five-loop of the QCD coupling constant $\alpha_s(Q)$ at zero temperature, as well as two- and three-hard thermal loop perturbation, have been calculated and results have been compared with available experimental data. There exists an appropriate agreement between theory and experiment at high $Q$'s. But at low energies, both five-loop zero temperature and three-loop finite temperature (as best approximations) cannot predict correct values for   $\alpha_s(Q)$. This means that considering temperature and chemical potential at low energies cannot fully compensate the deviation from experimental data at zero temperature calculations. We have applied the q-nonextensive treatment of running coupling on two- and three-loop perturbation in field theory at finite temperature. Our results show that, calculated values of strong coupling in a non-extensive regime at finite temperature can successfully generate experimental data by $q=1.2412$ (with $\chi^2=12.24$) at two-loop thermal QCD approximation. The non-extensive parameter for the three-loop approximation is $q=1.3181$ with a total variance $\chi^2=11.93$. Obtained non-extensive values clearly show a considerable deviation from standard Maxwellian statistics.\

Therefore we can conclude that, nonextensivity is a very important issue at least for low energy values of $\alpha_s(Q)$ in QCD calculations. Nonextensivity of hadronic systems, which is reflected in the strong coupling constant, certainly creates interesting effects especially in statistical features of QCD problems, like calculating parton distribution functions (PDFs), jet evolution, spin statistics and many other research interests in the field.

\end{document}